\documentclass{ws-ijmpa}
\usepackage[super,compress]{cite}
\usepackage{epsfig}
\usepackage{color}
\begin{document}

\markboth{Hemwati Nandan, Akhilesh Ranjan}
{Regge Trajectories of Exotic Hadrons in the Flux Tube Model}

%
\catchline{}{}{}{}{}
%

\title{Regge Trajectories of Exotic Hadrons in the Flux Tube Model}

\author{\footnotesize Hemwati Nandan}
\address{Department of Physics, Gurukula Kangri Vishwavidyalaya,
249 404 Haridwar, INDIA \\
hnandan@iucaa.ernet.in}

\author{Akhilesh Ranjan}
\address{Department of Physics, Manipal Institute of Technology, Manipal
University, Manipal, 576 104 Udupi, INDIA\\
akhileshxyz@yahoo.com}

\maketitle

\begin{history}
\end{history}

\begin{abstract}
We have investigated the Regge trajectories of exotic hadrons  by considering 
different possible pentaquark configurations with finite quark mass in the  
flux tube model. Significant deviation is observed in the linear behavior of 
the Regge trajectories for pentaquark systems in view of the universal value 
of the Regge slope parameter for hadrons. The modified Regge trajectories 
are also compared with the available experimental and lattice data. It is 
observed that the non-linear Regge trajectories of such pentaqurk systems can 
be well described by the relativistic corrections in view of the current quark 
masses and the high rotational speed of the quarks at the end of flux tube 
structure. 

\end{abstract}

\ccode{PACS numbers: 11.55Jy, 14.20.-c, 14.40.-n.}

\section{Introduction}	

According to quark spectrum and color confinement different kind of color 
singlet quark configurations can exist in nature \cite{guidry}. Some of these 
confugurations are exotic hadrons and the presence of such exotic hadrons in 
nature is believed to provide useful insights into the largely unknown dynamics 
of quark-gluon dynamics mainly with respect to the confinement and other 
closely related aspects. In general, the linear Regge trajectories of hadrons 
admit a linear confinement potential \cite{guidry,cheng} and therefore, it is 
also important to investigate the behavior of such trajectories for the case 
of exotic hadrons. Therefore, in order to further understand the quark-gluon 
interactions and properties of quark matter, it is useful to investigate the 
Regge trajectories of the exotic hadrons. It is worth noticing that many of 
the resonance states observed at Belle, Babar and CLEO experiments 
are quite difficult to interpret in terms of conventional mesonic and baryonic 
states \cite{chen,wang} and therefore, some other color singlet 
quark combinations are proposed viz. exotic mesons, 
tetraquarks, hybrid mesons etc. It is believed that these particles 
may actually be hybrids or 
hadron molecules and some of the exotic hadrons also belong to pentaquark 
family which consists four quarks and one antiquark. The primary aim here is 
to further explore the properties of such pentaquark configurations. The 
theoretical prediction of such systems which are heavy and short lived states 
was first given by M. Jezabek and  M. Praszalowicz \cite{jezabek} in 1987.  
Initially, such states were observed as exotic baryons produced in different 
processes \cite{exoticbaryons}, but with the improved experimental facilities, 
the experimental observation of these particles was first reported by Diakonov, 
Petrov, and Polyakov \cite{diakonov} in 1997. Since then many experiments have 
confirmed the existence of different pentaquark systems \cite{exp,sonia,liu} 
and their properties are discussed in detail based on lattice calculations 
\cite{jahn}. The analysis of many experimental data still need more careful 
attention to prove the existence of pentaquark configurations 
\cite{torres,hicks} however, there are number of investigations to confirm 
the existence of such states including the recent one at CERN and DIANA 
collaborations 
\cite{jahn,charmed-pentaquarks,cern,diana}. One may also notice that there are 
evidences for charmed  pentaquarks  \cite{charmed-pentaquarks}  in addition to 
the first observed pentaquark $\Theta^{+}$. \\

The properties of pentaquarks have been studied by using different models 
\cite{stancu,ping}. The flux tube model has been very successful in predicting 
the classical mass as well as the angular momentum  of hadrons 
\cite{regge-1,regge-2,chew,hothi,inopin-2,sharov-1,hnregge,corr-rt,hnandan} 
and it has 
also been used to study the Regge trajectories of pentaquark configurations 
with diquark-antiquark-diquark clustering in rotating flux tube at different 
orbital excitations \cite{martemyanov}.

In the present work,  we consider the flux tube model to extend our earlier 
work \cite{corr-rt} on Hadronic Regge trajectories to the case of pentaquarks 
system. A generalized formulation of the Regge trajectories of pentaquarks is 
developed by incorporating the mass of quarks. One-four and two-three 
quark/antiquark clustering is considered at the end of flux tube. It is found 
that the Regge trajectories of pentaquarks are highly non-linear and the 
results obtained are also compared with the experimental and lattice data. 

\section{The Classical Mass and Angular Momentum}
In the flux tube model of hadrons, with the massless quarks lying at the end of 
the string, it is assumed that the endpoints of the string rotate with the 
speed of light. The mass of the hadron then emerges as the potential energy 
due to the string tension. Let length of the string is $l$ and string tension 
is $\sigma$. If the string is rotating about the mid point, the mass of hadron 
in the natural  system of units (i.e., $\hbar=1$ and $c=1$) is given as 
\cite{cheng}
\begin{eqnarray}
\label{eqn-1}
M=2\int^{\frac{l}{2}}_0\frac{\sigma\,dr}{\sqrt{1-v^2}},
\end{eqnarray}
The angular momentum of the hadron is however, given as 
\begin{eqnarray}
\label{eqn-2}
J=2\int^{\frac{l}{2}}_0\frac{\sigma vr\,dr}{\sqrt{1-v^2}},
\end{eqnarray}
The inter-relation between $J$ and $M$ reads as
\begin{eqnarray}
\label{eqn-3}
J=\alpha' M^2+\alpha_{0} 
\end{eqnarray} 
where $\alpha_0$ is a constant which is occuring due to the intrinsic spin of 
quarks and the Regge slope parameter $\alpha'$ is given by 
$\frac{1}{2\pi \sigma}$. The linear potential between quarks is given 
by $V(r)=\sigma r$ where $r$ is the sparation between quarks. \\

\begin{figure}[ph]
\centerline{\psfig{file=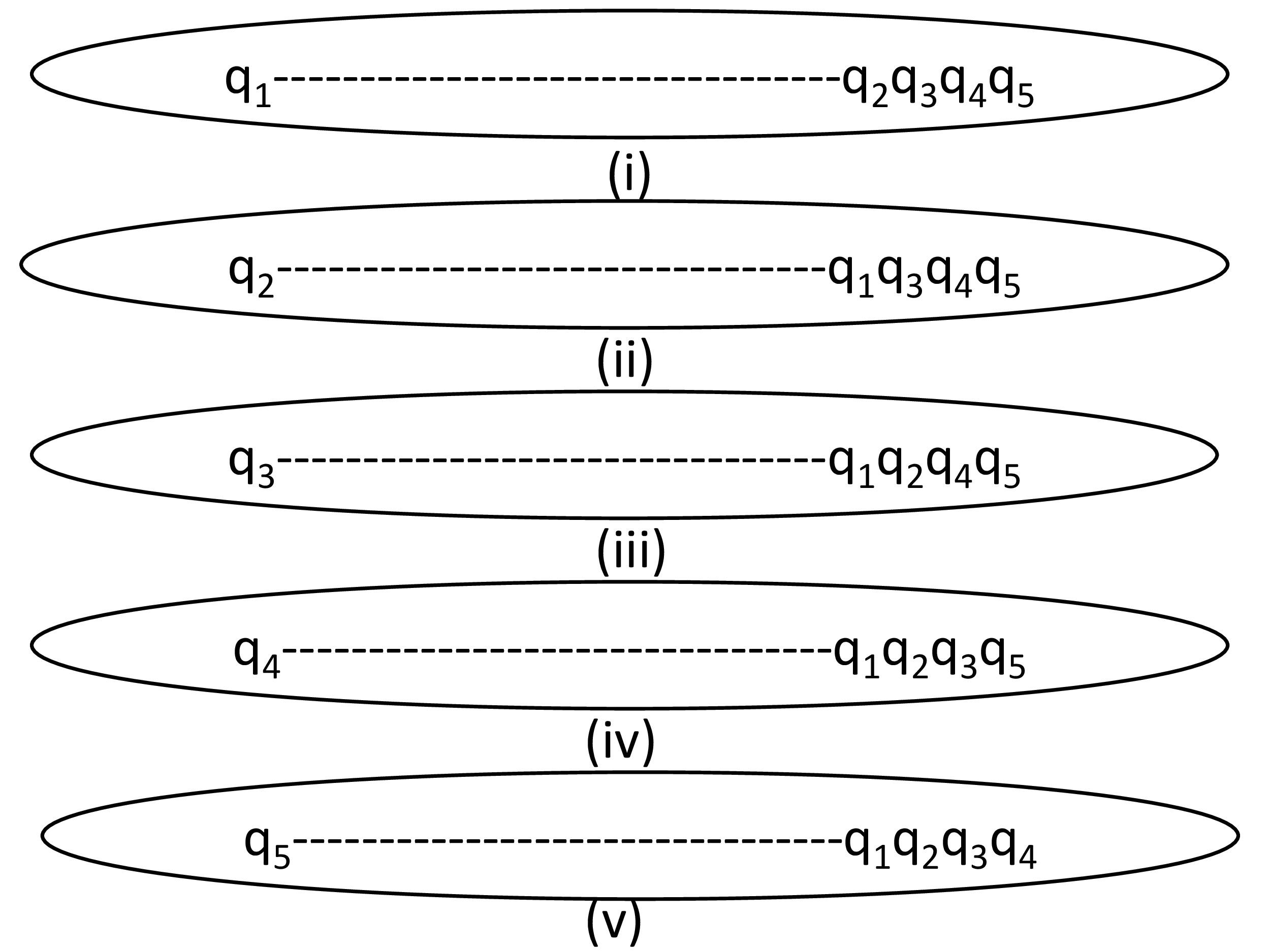,height=8.0 cm, width=6.0 cm}}
\vspace*{-0mm}
\caption{ Different configurations of pentaquarks with one quark at one end of the string.\protect\label{pqca}}
\end{figure}

\begin{figure}[ph]
\centerline{\psfig{file=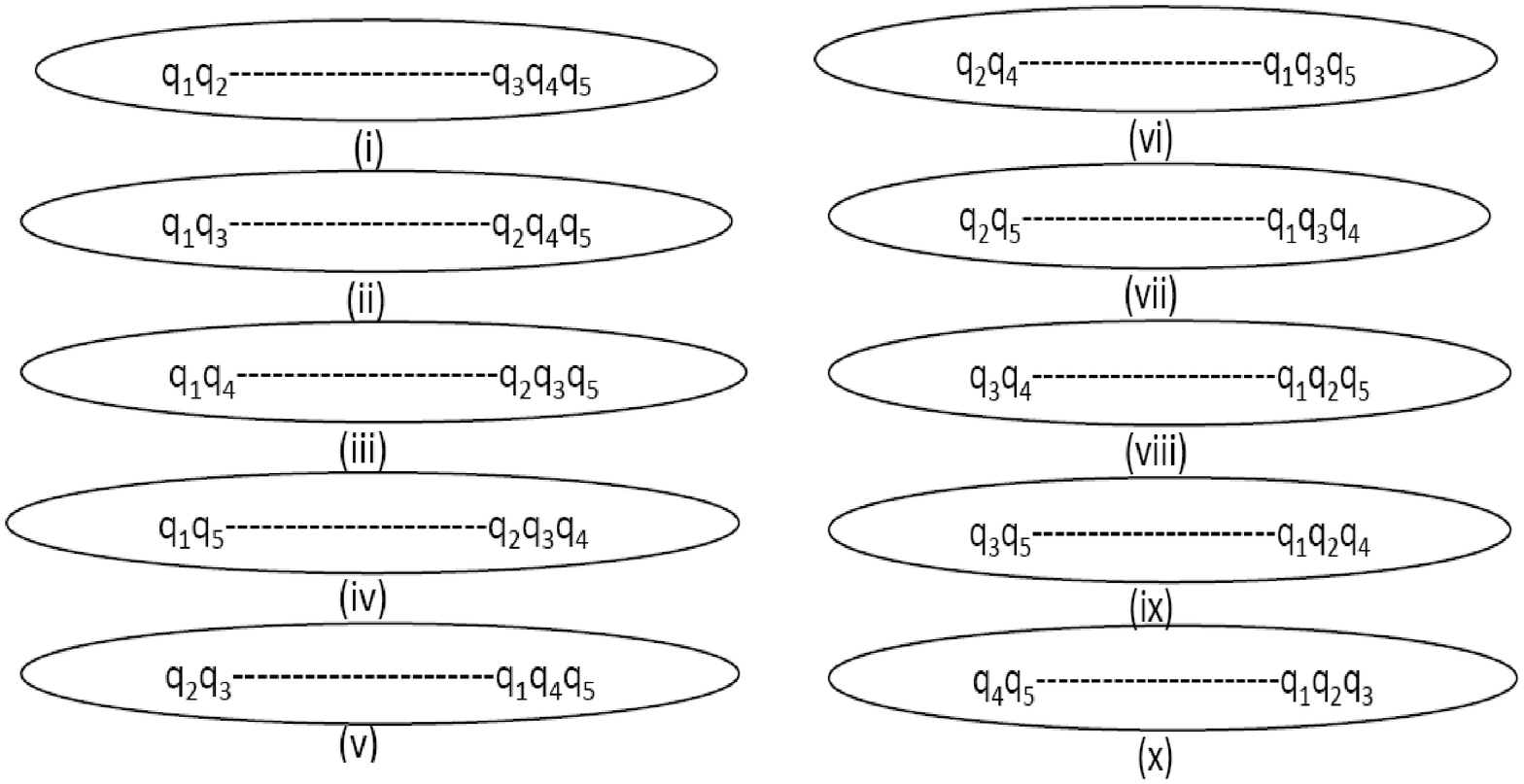,height=10.0 cm, width=7.0 cm}}
\vspace*{-15.0mm}
\caption{Different configurations of pentaquarks with two quarks at one end of the string.\protect\label{pqcbt}}
\end{figure}

Recently, we  have studied the Regge trajectories of hadrons by using the 
massive qaurks \cite{corr-rt}. Here, we consider different possible 
configurations of pentaquark systems as presented in  Fig\ref{pqca}  and  
Fig\ref{pqcbt} and if one extends the formulation in \cite{corr-rt} for  
pentaquarks, the modified mass expression for the pentaquark for configuration 
(i) as in Fig\ref{pqca} reads as
\begin{eqnarray} 
\label{pqc1m1}
M_{1i}&=&\frac{ \sigma (M-m_1)l}{fM}\left(\int_{0}^{f}\frac{\,dv}{\sqrt{1-v^2}}+\int_{0}^{\frac{m_1}{(M-m_1)}f}\frac{\,dv}{\sqrt{1-v^2}} \right)\nonumber\\
&+&\gamma_1m_1+\gamma_2(M-m_1)
\end{eqnarray} 
where $M=m_1+m_2+m_3+m_4+m_5$, $\gamma_1=\frac{1}{\sqrt{1-f^2}}$ and
$\gamma_2=\frac{1}{\sqrt{1-\frac{m^2_1f^2}{(M-m_1)^2}}}$.  Here $f$ is the 
fractional rotational speed (actual speed is $fc$) of the endpoint on string at 
position $r$ from the mid point of the string. The first two terms in the 
above expression show the flux tube contribution while the other two terms 
are relativistic masses of quarks. The Eq (\ref{pqc1m1}) after integration can 
be rewritten as follows,
\begin{eqnarray} 
\label{pqc1m2}M_{1i}&=&\frac{ \sigma (M-m_1)l}{fM}\left(sin^{-1}f+sin^{-1}\frac{m_1f}{(M-m_1)}\right)\nonumber\\
&+&\gamma_1m_1+\gamma_2(M-m_1)  
\end{eqnarray} 
The expression for the angular momentum of pentaquark ($J_{1i}$) is given by 
\begin{eqnarray}
\label{pqc1j1} 
J_{1i}&=&\frac{ \sigma  (M-m_1)^2l^2}{f^2M^2}\left(\int_{0}^{f}\frac{v^2\,dv}{\sqrt{1-v^2}}+\int_{0}^{\frac{m_1}{(M-m_1)}f}\frac{\,dv}{\sqrt{1-v^2}} \right)\nonumber\\
&+&\frac{m_1fl}{M}\{\gamma_1(M-m_1)+\gamma_2m_1\}  
\end{eqnarray} 
where the first two terms represent the angular momentum generated due to 
rotation of string and the other two terms represent the angular momentum 
generated due to motion of quarks. After, integration, the Eq (\ref{pqc1j1}) 
leads to the following form 
\begin{eqnarray} 
\label{pqc1j2}
J_{1i}&=&\frac{ \sigma (M-m_1)^2l^2}{f^2M^2}\left(\frac{1}{2}sin^{-1}f-\frac{f}{2}\sqrt{1-f^2}+\frac{1}{2}sin^{-1}\frac{m_1f}{M-m_1}\right.\nonumber\\ 
&-&\left.\frac{m_1f}{2(M-m_1)}\sqrt{1-\frac{f^2m_1^2}{(M-m_1)^2}}\right)\nonumber\\
&+&\frac{m_1fl}{M}(\gamma_1(M-m_1)+\gamma_2m_1)
\end{eqnarray} 
One can easily notice that the above expressions are directly dependent on 
$(M-m_1)$ and the configuration containing the heaviest quark at one end will
have least contribution among all other configurations.  Similarly for 
configuration (i) as presented in Fig\ref{pqcbt}, the expression for mass 
reads as
\begin{eqnarray} 
\label{pqc2m1}
M_{2i}&=&\frac{ \sigma (M-m_1-m_2)l}{fM}\left(\int_{0}^{f}\frac{\,dv}{\sqrt{1-v^2}}+\int_{0}^{\frac{m_1+m_2}{M-m_1-m_2}f}\frac{\,dv}{\sqrt{1-v^2}} \right)\nonumber\\
&+&\gamma_3(m_1+m_2)+\gamma_4(M-m_1-m_2)
\end{eqnarray} 
where $\gamma_3=\frac{1}{\sqrt{1-f^2}}$ and
$\gamma_4=\frac{1}{\sqrt{1-\frac{(m_1+m_2)^2f^2}{(M-m_1-m_2)^2}}}$. The string 
length is assumed same for all the configurations. The Eq (\ref{pqc2m1}), can be 
rewritten as follows after integration 
\begin{eqnarray} 
\label{pqc2m2}
M_{2i}&=&\frac{ \sigma (M-m_1-m_2)l}{fM}\left(sin^{-1}f+sin^{-1}\frac{(m_1+m_2)f}{(M-m_1-m_2)}\right)\nonumber\\
&+&\gamma_3(m_1+m_2)+\gamma_4(M-m_1-m_2)  
\end{eqnarray} 
The modified angular momentum of pentaquark ($J_{2i}$) is given by 
\begin{eqnarray}
\label{pqc2j1} 
J_{2i}&=&\frac{ \sigma (M-m_1-m_2)^2l^2}{f^2M^2}\left(\int_{0}^{f}\frac{v^2\,dv}{\sqrt{1-v^2}}+\int_{0}^{\frac{m_1+m_2}{M-m_1-m_2}f}\frac{\,dv}{\sqrt{1-v^2}} \right)\nonumber\\
&+&\frac{(m_1+m_2)fl}{M}\{\gamma_3(M-m_1-m_2)+\gamma_4(m_1+m_2)\}  
\end{eqnarray} 
which can be simplified as follows after integration 
\begin{eqnarray} 
\label{pqc2j2}
J_{2i}&=&\frac{ \sigma (M-m_1-m_2)^2l^2}{f^2M^2}\left(\frac{1}{2}sin^{-1}f-\frac{f}{2}\sqrt{1-f^2}+\frac{1}{2}sin^{-1}\frac{(m_1+m_2)f}{M-m_1-m_2}\right.\nonumber\\ 
&-&\left.\frac{(m_1+m_2)f}{2(M-m_1-m_2)}\sqrt{1-\frac{f^2(m_1+m_2)^2}{(M-m_1-m_2)^2}}\right)\nonumber\\
&+&\frac{(m_1+m_2)fl}{M}(\gamma_3(M-m_1-m_2)+\gamma_4(m_1+m_2))
\end{eqnarray} 
In Fig\ref{pqca} and Fig\ref{pqcbt}, there are fifteen configurations in total 
which are equally probable. Therefore, the mass and angular momentum of the 
pentaquark should be averaged over all such configurations. The expressions of 
mass and angular momentum for configurations `1i' and `2i' are functions of 
$sin^{-1}\frac{m_1f}{M-m_1}$ and $sin^{-1}\frac{(m_1+m_2)f}{M-m_1-m_2}$ 
respectively. Since $sin\theta\le 1$ so $f\le\frac{M-m_1}{m_1}$ and 
$f\le\frac{M-m_1-m_2}{m_1+m_2}$. Further, according to the special theory of 
relativity $f\le1$. Such conditions would also arise for all the configurations 
which shall satisfy simultaneously. 

\section{Summary and Conclusions}
 We have considered the mass of up, down, strange and charm quarks as $m_u=2.3$ 
MeV, $m_d=4.8$ MeV, $m_s=95$ MeV, and $m_c=1275$ MeV respectively, and the 
string tension\cite{pdg} $\sigma=0.2 \,GeV^{2}$. The length of string should 
change for different quark masses in order to maintain the string tension such 
that the angular momentum remains constant. Therefore, the length of the string 
for different pentaquark configurations is taken different as given in table 1. 
 In the last column of the table the contribution of the pentaquark 
configuration when the heaviest quark is at one end of the string is shown. It 
is, therefore, clear that such configuration results are in close agreement 
with the actual ones.     

\begin{figure}[ph]
\centerline{\psfig{file=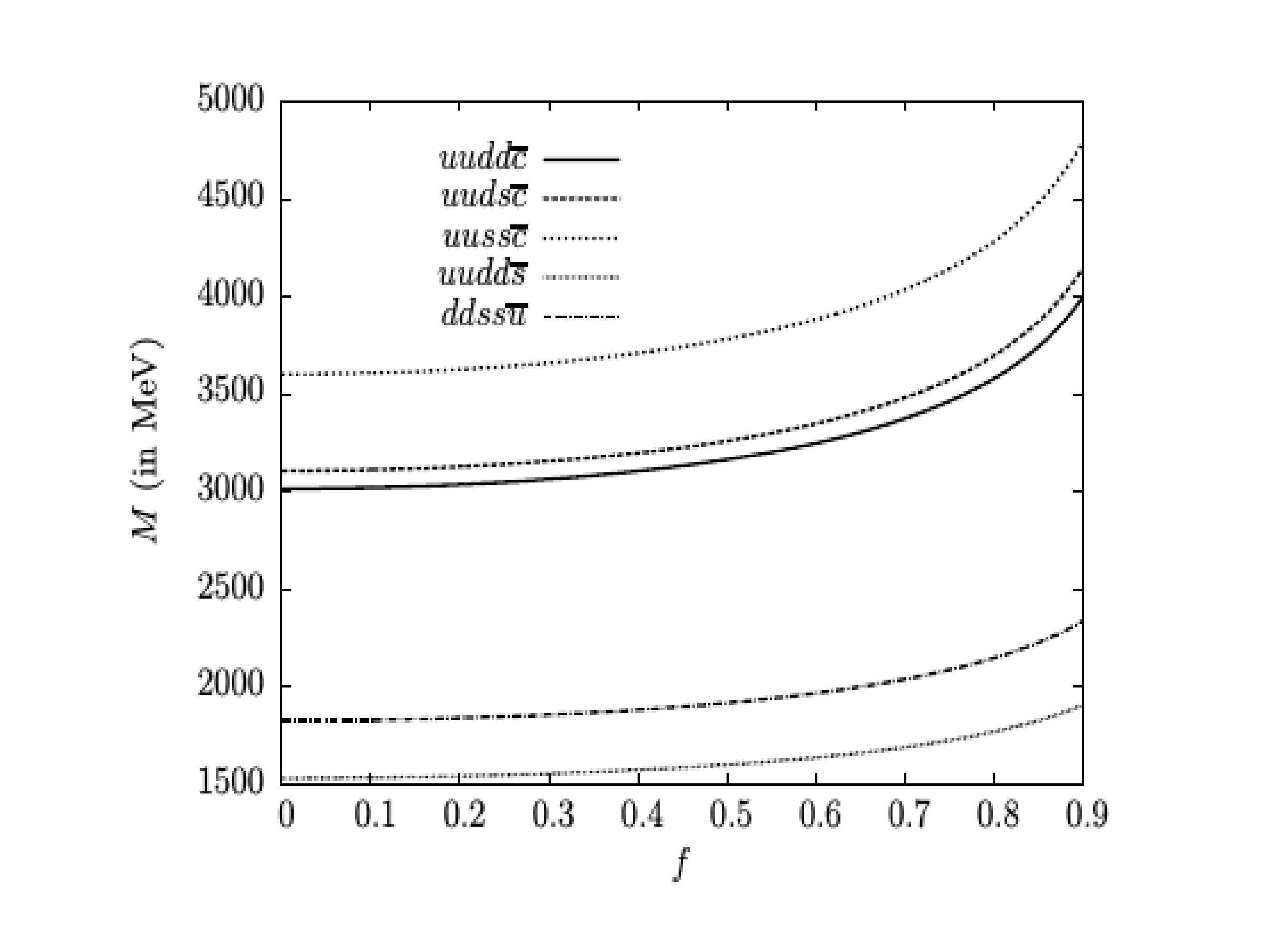,height=7.0 cm, width=12.0 cm}}
\vspace*{-5.0mm}
\caption{Mass (M) variation of different pentaquarks with variation in speed 
(f) of lighter end of pentaquark string.\protect\label{mod-f-m}}
\end{figure}

The mass variation of pentaquarks with the rotational speed is presented in 
Fig\ref{mod-f-m} which indicates almost the same pattern for all the 
pentaquark configurations. For lower speed, it is almost linear but for higher 
speed, it shows a highly non-linear behavior. The seperation between each 
curve depends upon the current quark mass difference of pentaquarks. One may 
notice that there is a large gap between charm pentaquarks and non-charm 
pentaquarks as shown in Fig\ref{mod-f-m}. For the purpose of computation, the 
rotational speed of the low mass string end is taken as $fc<1$ ($c=1$) and mass 
as well as angular momenta are then calculated for different values of $f$. 
It is observed that for string length range 1.4 - 2.1 fm, our results as 
presented in table 1 are in good agreement with the experimental 
\cite{sonia,liu} and lattice \cite{stancu} data. It is interesting to note 
that as the mass of 
pentaquarks increases the string length also increases. But from simple 
classical analysis, it can be proved that if the angular momentum and string 
tension are constant then the string length should decrease with mass. 
According to the special theory of relativity, at constant force a light mass 
system will be more relativistic than a heavier one which can easily be 
visualised in our data. It is found that for light pentaquarks nearly 90$\%$ 
mass is coming due to the relativistic correction and for heavier pentaquarks 
it is nearly 50$\%$. 

\begin{figure}[ph]
\centerline{\psfig{file=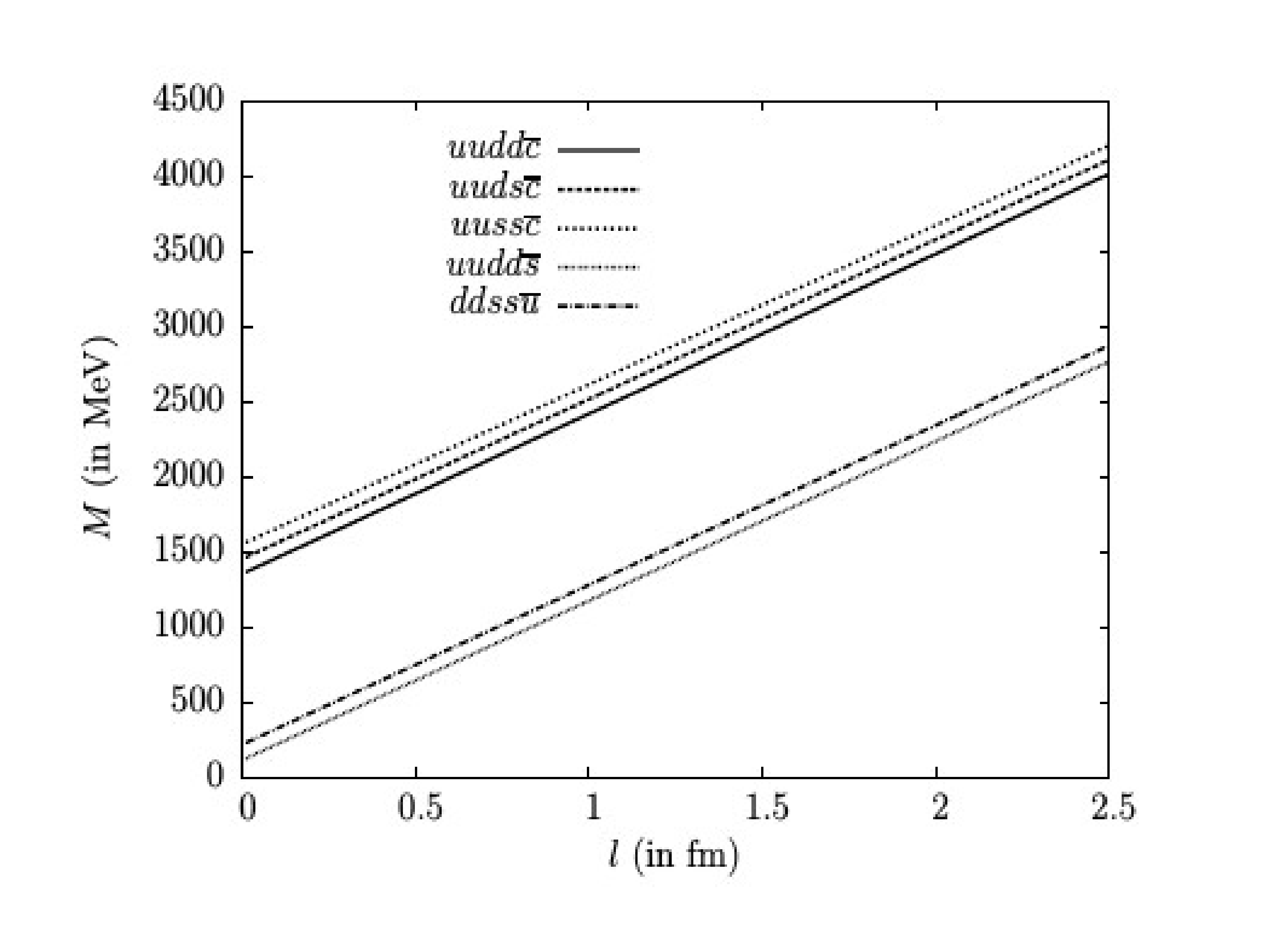,height=7.0 cm, width=12.0 cm}}
\vspace*{-5.0mm}
\caption{Mass (M) variation of different pentaquarks with variation in string 
length.\protect\label{l-m1}}
\end{figure}

Fig\ref{l-m1} shows the mass dependence of pentaquarks with 
string length. Here, we have considered the fractional rotational speed of 
lower mass end of pentaquark string as 0.5 which is an arbitrary number but 
seems to be reasonable for both the light and heavy pentaquarks. One can 
easily notice that the mass of the pentaquarks varies linearly with the 
string length. The intersection points on the mass axis denotes the 
contribution of the current quark mass on the actual pentaquark mass. One can 
infer the string lengths for different pentaquarks from this graph by 
comparing it with the experimental data.  

\begin{figure}[ph]
\centerline{\psfig{file=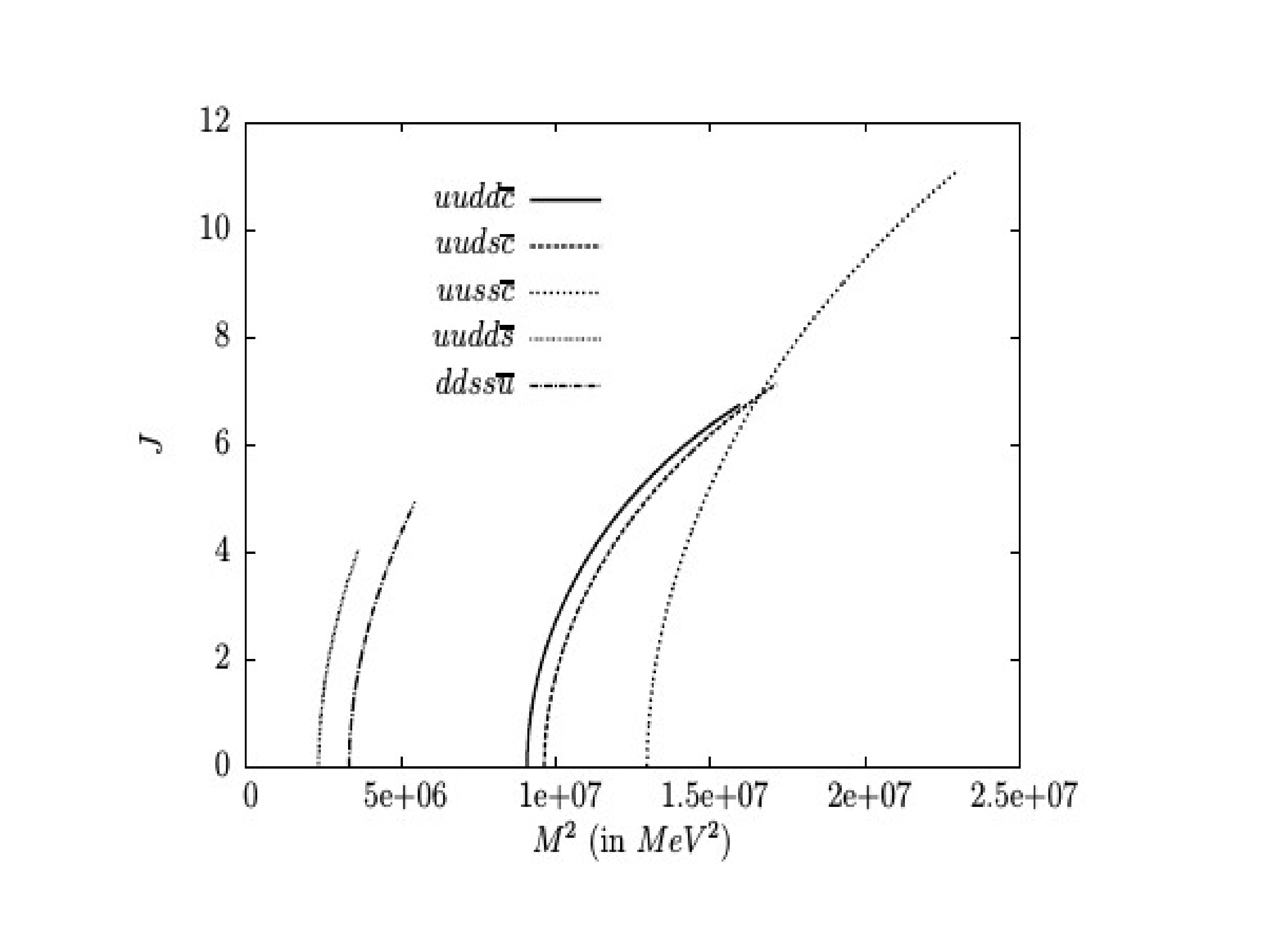,height=7.0 cm, width=12.0 cm}}
\vspace*{-5.0mm}
\caption{Regge trajectories for different pentaquarks.\protect\label{pq-rtmod}}
\end{figure}

In order to perform the calculation of Regge trajectories, the 
averaged expressions of angular momentum (J) and mass (M) over all 
configurations are considered and the Fig\ref{pq-rtmod} represents a graph 
between averaged angular momentum and square of mass. In order to get the 
desired expressions, first the relation between $J$ and $M$ for any single 
quark configuration of pentaquark is calculated and then the average angular 
momentum over all the configurations in terms of $M$ is obtained. One can 
easily notice that it will not be proportional to the square of averaged mass 
of pentaquark (over all the quark configurations). On following an analysis 
for baryons by Nandan and Ranjan \cite{corr-rt}, it can easily be shown that 
the Regge trajectory for the pentaquark will not be linear.
The non-linear Regge trajectories are found as depicted in Fig\ref{pq-rtmod} 
and it is interesting to note that for heavy pentaquarks the angular momentum 
spectrum is richer than for light pentaquarks. Two Regge trajectories which 
intersect each other indicates that two different pentaquarks can have same 
angular momentum as well as mass. From the expressions of angular momenta and 
mass, it appears that the steepness of the Regge trajectories will depend upon 
the current quark mass and string length. The steepness of the trajectories 
increases  with the increase in the 
string length, but for quarks' current mass it is not so apparent. In the Fig 
\ref{pq-rtmod}, string length for pentaquarks $uudd\overline{c}$ and 
$uuds\overline{c}$ are same but still they intersect each other which means 
that the steepness of the Regge trajectory decreases with decrease in the 
current quark mass. From Fig\ref{pq-rtmod}, one may also conclude that 
the steepness of Regge trajectories decreases with mass of pentaquarks. 
This discussion becomes particularly important when one replaces the slope 
of the linear portion of the Regge trajectories by effective string tension. 

Therefore, one may conclude that the Regge trajectories of pentaquarks is 
well described by the relativistic corrections with current quark masses. 
The mass of the pentaquark increases almost linearly with low rotational speed 
of the string but becomes highly non-linear at high rotational speed and also 
causes non-linearity in the Regge trajectories. If we wish to interpret the 
slope of the linear portion of the Regge trajectories as effective string 
tension then it will depend upon the string length and current quark mass. 
The interesting result we observed from our investigations is that two 
different pentaquarks can have equal mass as well as angular momentum. Still 
we need more experimental data to verify these results. In view of these 
results, one may expect the emergence of more/new pentaquarks from colliders
 in near future.

\begin{table}[ph] 
\label{table1} 
\tbl{Comparision of our results with other's works on pentaquarks.}
{\begin{tabular}{|l|l|l|l|l|l|l|}
\hline
S. No.&Pentaquark& Quark &Our results& Other's&&Charm\\
&&structure&Mass(J)&results&String&quark end\\
&&&in MeV&Mass (J)&length&contribution\\
&&&&in MeV&in fm&Mass(J)\\
&&&&\cite{stancu,sonia,liu}&&in MeV\\ 
\hline
1.&$\Theta^o_c$&uudd$\bar c$&3020($\frac{1}{2}$)&3099($\frac{1}{2}$)&1.7&3024($\frac{1}{2}$)\\
\hline
2.&$N^o_c$&uuds$\bar c$&3110($\frac{1}{2}$)&3180($\frac{1}{2}$)&1.7&3115(0.511)\\
\hline
3.&$\Xi^o_c$&uuss$\bar c$&3604($\frac{1}{2}$)&3650($\frac{1}{2}$)&2.1&3606(0.508)\\
\hline
4.&$\theta^+$&uudd$\bar s$&1537($\frac{1}{2}$)&1540($\frac{1}{2}$)&1.4&1538(0.448)\\
\hline
5.&$\Xi^{--}_c$&ddss$\bar u$&1883($\frac{3}{2}$)&1860($\frac{3}{2}$)&1.6&1885.5(1.227)\\
\hline
\end{tabular}}
\end{table}
 
\section*{Acknowledgements} 
One of the authors HN is thankful to the Department of Science and Technology 
(DST), New Delhi for the financial assistance through grant number  
SR/FTP/PS-31/2009. Authors are also thankful to the anonymous referees 
for their critical and valuable suggestions which helped in improving the 
presentation of this paper.


\begin{thebibliography}{0}

\bibitem{guidry}
Mike Guidry, 
{\it Gauge Field Theories An Introduction with Applications}, 
(Wiley-VCH Verlag GmbH $\&$ Co. KGaA, 1991).

\bibitem{cheng} 
Ta-Pei Cheng, Ling-Fong Li, 
{\it Gauge theory of elementary particle physics}, 
(Claredon Press, Oxford 1982).

\bibitem{chen}
Wei Chen, T. G. Steele, Shi-Lin Zhu,
Universe {\bf 2} 13(2014), arXiv:1403:7457v1 [hep-ph].

\bibitem{wang}
X. L. Wang {\it et al} (Belle), 
Phys. Rev. Lett. {\bf 99}, 142002(2007);
M. Ablikim {\it et al} (BESIII Collaboration), 
Phys. Rev. Lett. {\bf 112} 132001(2014), arXiv:1308.2760 [hep-ex];
W. Chen, T. Steele, M. L. Du, and S. L. Zhu,
Eur. Phys. J C {\bf 74} 2773(2014); 
C. Y. Cui, Y. L. Liu, and M. Q. Huang,
Eur. Phys. J. C {\bf 73} 2661(2013), arXiv:1308.3625 [hep-ph];
B. Aubert {\it et al} (BABAR), 
Phys. Rev. Lett. {\bf 95} 142001(2005), arXiv:0506.081 [hep-ex];
Q. He {\it et al} (CLEO), 
Phys. Rev. D {\bf 74}   091104(2006), arXiv:0611.021 [hep-ex];
C. Z. Yuan {\it et al} (Belle), 
Phys. Rev. Lett. {\bf 99} 182004(2007). 

\bibitem{jezabek}
M. Jezabek, M. Praszalowicz, 
{\it Proceedings of the Workshop on Skyrmions and Anomalies}, Krakow, Poland, 
1987. World Scientific. p. 112.


\bibitem{exoticbaryons}
T. Takano {\it et al}, 
Phys. Rev. Lett. {\bf 91} 012002(2003), arXiv:hep-ex/0301020v2;
V. V. Barmin {\it et al},
Phys. Atom Nucl. {\bf 66} 1715(2003), arXiv:hep-ex/0304040v4;
S. Stepanyan {\it et al}
Phys. Rev. Lett. {\bf 91} 252001(2003), arXiv:hep-ex/0307018v4;
A. E. Asratyan, A. G. Dolgolenko, and M .A. Kubantsev,
Phys. Atom. Nucl. {\bf 67} 682(2004), arXiv:hep-ex/0309042v3;
V. Kubarovsky {\it et al}, 
Phys. Rev. Lett. {\bf 92} 032001(2004), Erratum-ibid.92:049902,2004, 
arXiv:hep-ex/0311046v3;
A. Airapetian {\it et al},
Phys. Lett. B {\bf 585} 213(2004), arXiv:hep-ex/0312044v2;
A. Aleev {\it et al}, 
Phys. Atom. Nucl. {\bf 68} 974(2005), arXiv:hep-ex/0401024v5;
Zeus Collaboration,
Phys. Lett. B {\bf 591} 7(2004), arXiv:hep-ex/0403051v2.

\bibitem{diakonov}
D. Diakonov, V. Petrov, and M. Polyakov,
Z. Phys. A {\bf 359} 305(1997), arXiv:hep-ph/9703373.

\bibitem{exp}
J. Barth {\it et al},
Phys. Lett. B {\bf 572} 127(2003), arXiv:hep-ex/0307083v4;
Daniel S. Carman,
Eur. Phys. J. A {\bf 24} S1 15(2005), arXiv:hep-ex/0412074v1;
Kenneth H. Hicks,
Prog. Part. Nucl. Phys. {\bf 55} 647(2005), arXiv:hep-ex/0504027v2;
Michael Danilov and Roman Mizuk, 
Phys. Atom. Nucl. {\bf 71} vol 4 605(2008), arXiv:0704.3531v2 [hep-ex].

\bibitem{sonia}
Sonia Kabana,
J. Phys. G {\bf 31} S1155(2005), arXiv:hep-ex/0503019v1;

\bibitem{liu}
Tianbo Liu, Yajun Mao, and Bo-Qiang Ma,
arXiv:1403.4455v1 [hep-ex].

\bibitem{jahn} 
O. Jahn, J. W. Negele, and D. Sigaev,
Proceedings of Science LAT2005 069(2006), arXiv:hep-lat/0509102v1. 

\bibitem{torres}
A. Martinez Torres and E. Oset,
{\it Proceeding for the CHIRAL 10 workshop}, Valencia (Spain), 2010, 
arXiv:1012.2967v1 [nucl-th]. 

\bibitem{hicks}
Ken Hicks,
arXiv:hep-ph/0703004v3.

\bibitem{charmed-pentaquarks} 
S. M. Gerasyuta, V. I. Kochkir, and Xiang Liu,
arXiv:1407.2702v1 [hep-ph]. 

\bibitem{cern}
http://home.web.cern.ch/about/updates/2015/07/discovery-new-class-particles-lhc,
LHCb Collaboration,
arXiv:hep-exp/1507.03414.

\bibitem{diana}
DIANA Collaboration,
arXiv:hep-exp/1507.06001.

\bibitem{stancu} 
Fl. Stancu, 
Int. J. Mod. Phys. A {\bf 20} 209(2005), arXiv:hep-ph/0408042v1. 

\bibitem{ping} 
Jialun Ping {\it et al}, 
Phys. Rev. C {\bf 77} 025201(2008), arXiv:0802.2891v1. 

\bibitem{regge-1} 
T. Regge, 
{\it Nuovo Cimento} {\bf 14} 951(1959). 

\bibitem{regge-2} 
T. Regge, 
{\it Nuovo Cimento} {\bf 18} 947(1960). 

\bibitem{chew} 
G. F. Chew and S. C. Frautschi, 
{\it Phys. Rev. Lett.} {\bf 8} 41(1962). 

\bibitem{hothi} 
Shuchi Bisht, Navjot Hothi, and Gaurav Bhakuni, 
{\it EJTP} {\bf 7} 299(2010). 

\bibitem{inopin-2} 
A. Inopin and G. S. Sharov, 
{\it Phys. Rev. D} {\bf 63} 054023(2001), arXiv:hep-ph/9905499. 

\bibitem{sharov-1} 
G. S. Sharov, 
{\it Phys. Rev. D} {\bf 62} 094015(2000), arXiv:hep:ph/0004003. 




\bibitem{hnregge}
{\it Proceedings of the DAE Symp. on Nucl. Phys.} {56} (2011) 706;
{ibid  56 (2011) 708}.

\bibitem{corr-rt}
Akhilesh Ranjan, Hemwati Nandan,
Mod. Phys. Lett. A {\bf 27} vol. 8 1250047(2012).

\bibitem{hnandan}
Hemwati Nandan, T. Anna and H.C. Chandola,
Euro Phys. Lett. {\bf  67(5)} 746 (2004);
Hemwati Nandan, H. C. Chandola 
and H. Dehnen, Int. J. Theor. Phys. {\bf 44(4)} (2004) 457 
and references therein; 
Hemwati Nandan, AIP Conference Proceedings {\bf 93} 17 (2007).

\bibitem{martemyanov}
B. V. Martemyanov {\it et al},
Phys. Rev. D {\bf 71} 017502(2005), arXiv:hep-ph/0502021v1. 

\bibitem{pdg}
J. Beringer {\it et al} (Particle Data Group), Phys. Rev. D {\bf 86}, 
010001(2012).

\end{thebibliography}
\end{document}